# Advancing Remote and Continuous Cardiovascular Patient Monitoring through a Novel and Resource-efficient IoT-Driven Framework

**Sanam Nayab[1], Sohail Raza Chohan[2], Aqsa Jameel[3], Syed Rehan Shah[1], Syed Ahsan Masud Zaidi[4], Aditya Nath Jha[4,], and Kamran Siddique[5]**

[1]Department of Computer Science, National University of Modern Languages, Islamabad 44000, Pakistan (sanam.nayab@numl.edu.pk) (rehan.shah@numl.edu.pk)

[2]Department of Computing & Emerging Technologies Emerson University Multan (sohail@eum.edu.pk)

[3]Department of Computer Science, The Women University, Multan 60000, Pakistan (aqsa.6023@wum.edu.pk)

[4]Department of Computer Science, Kansas State University, Manhattan, KS 66506, United States (ahsanzaidi@ksu.edu)

[4]Department of Physics, Kansas State University, Manhattan, KS 66506, United States (ahsanzaidi@ksu.edu), (adityajha@ksu.edu)

[5]Department of Computer Science and Engineering, University of Alaska Anchorage, Anchorage, United States (ksiddique@alaska.edu)

*Correspondence:* (rehan.shah@numl.edu.pk)

***Abstract*:** Cardiovascular diseases are a leading cause of fatalities worldwide, often occurring suddenly with limited time for intervention. Current healthcare monitoring systems for cardiac patients rely heavily on hospitalization, which can be impractical for continuous monitoring. This paper presents a novel IoT-based solution for remote, real-time tracking of critical cardiac metrics, addressing the pressing need for accessible and continuous healthcare, particularly for the aging population in Pakistan. The proposed IoT kit measures essential parameters such as body temperature, heart rate (HR), blood pressure (BP), oxygen saturation (SPO2), and electrocardiography (ECG).

A key innovation of the system is its integration with a cloud-based application, enabling constant remote monitoring and incorporating an alarm mechanism to alert medical professionals for timely intervention, reducing the risk of catastrophic incidents. The system was tested in a clinical environment with 20 participants, demonstrating results closely aligned with those obtained using standard medical devices. The findings validate the system's potential for reliable remote monitoring, offering a significant step forward in proactive cardiac healthcare management. This novel approach combines IoT technology with cloud-based applications to provide a cost-effective and efficient solution for reducing unexpected fatalities among cardiac patients.

1. Introduction

Globally, cardiovascular diseases (CVDs) account for 17.9 million deaths annually, resulting in the leading cause of mortality [1]. Arrhythmia is the primary cause of cardiac arrest and irregular cardiac impulse rhythm [2]. It can be elucidated in three distinct streams: bradycardia, tachycardia, and premature heartbeat. Based on 2024 healthcare data validated by the World Health Organization [3], arrhythmia is the primary reason for mortalities in at least 15% of cases. Cardiovascular disorders account for over 80% of sudden fatalities. The primary cause of acute stroke is atrial fibrillation.

Moreover, ventricular tachycardia, as opposed to cardiac arrest, is the primary cause of shock, which is the cessation of regular cardiac activity followed by abrupt death. With the advancement of technology, we must go from a passive to a ubiquitous condition to combat various heart illnesses. The most essential prerequisite for the system is real-time, continuous monitoring. These days, non-invasive sensors [4] and routine monitoring of vitally significant heart parameters are critical to the system's functioning properly. The primary goal of continuous cardiac monitoring is to provide a round-the-clock monitoring service. However, the main obstacles to establishing such a system in developing nations like Pakistan are inadequate infrastructure and resource constraints. Millions of electronic devices worldwide are linked to the internet via the IoT, and they all transmit and receive data using software-defined networks and cutting-edge network routing techniques [5-9].

The Internet of Things (IoT) is the way of the future [10], allowing devices, sensors, gadgets, cars, and other "things" to work together. Internet of Things (IoT) technology facilitates and monitors critical human processes regardless of where they are or what they are doing. IoT is a strategy that gives many people access to the internet's advantages at a very affordable cost and with the least effort. The system guarantees that it will continuously track and regulate the physiological symptoms [11] of cardiac patients, including temperature, blood pressure (BP), heart rate, SPO2, ECG [12], and related environmental indicators. The Internet of Things (IoT)-based cardiac care system is a powerful customized model that satisfies widespread healthcare safety and scientific requirements for heart problems and disorders.

One of the fastest-growing industries in the Internet of Things (IoT) in recent years is healthcare. To advance cardiovascular healthcare, which is undoubtedly a significant contributing factor in unexpected fatalities, a change from a passive to a widespread and omnipresent healthcare paradigm is required. A typical cardiac patient in our local hospitals faces several difficulties, such as waiting in lengthy lines, receiving expensive medical care, traveling from a distant location to see medical consultants regularly, and dealing with situations where the patient's condition is critical and life-threatening and necessitates immediate rescue and ongoing observation. The limited availability of clinical facilities in rural places exacerbates the problem. Figure 1 represents the IoT-based healthcare unit.

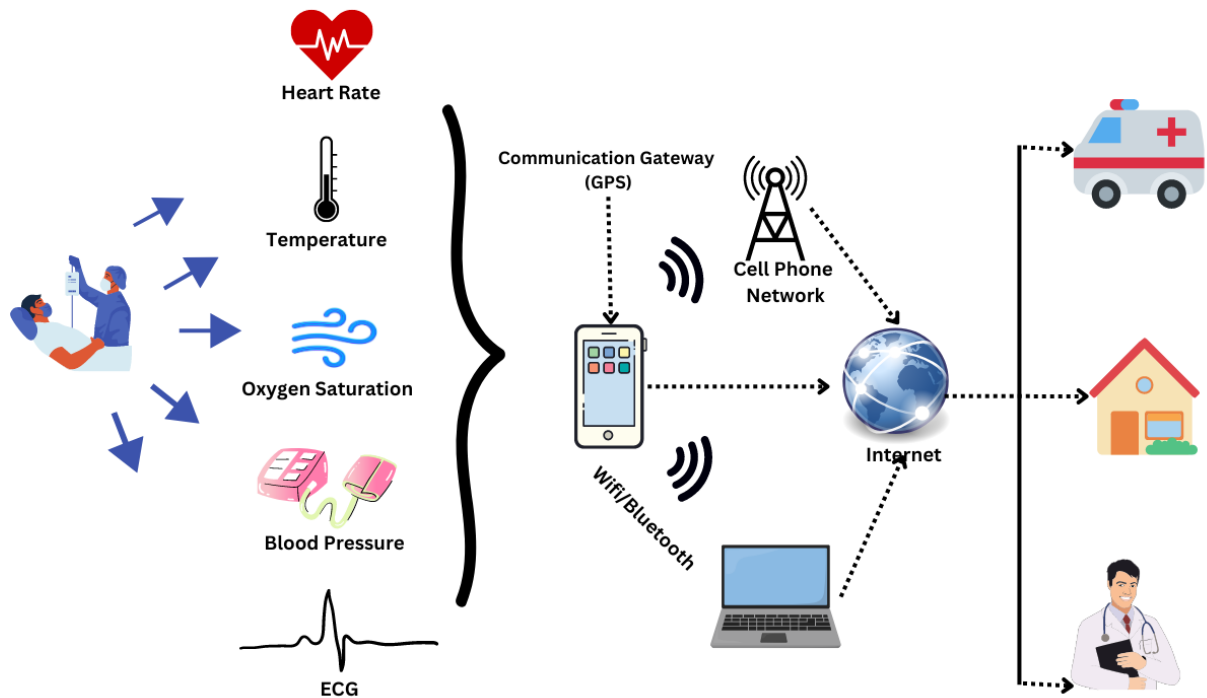

**Figure 1.** IoT-based Healthcare Unit

With the above in consideration, our study goals are:

- To propose a cloud-based IoT solution for a cardiovascular monitoring system that can help the doctor monitor cardiac patients at any time and place.
- To develop an IoT-based continuous cardiac healthcare system that can constantly monitor heart patients.
- Implement an IoT-based remote and continuous monitoring system that must include all cardiac patients' primary and vital parameters.
- To provide a continuous and remote solution with all the vital signs monitoring of the cardiac patients of Pakistan, which has yet to be done.

The novelty of this paper lies in its innovative use of cloud-based IoT technology for a wide-ranging and continuous cardiovascular monitoring system that considers all the essential signs of cardiac patients. This approach is implemented for cardiac patients in Pakistan, filling a noteworthy gap that has yet to be addressed.

Section 2 provides a compressive review of the relevant literature. Section 3 detailed about the proposed IoT-based Remote and Continuous Cardiovascular Patient Monitoring System. Section 4 presents the results and discussion of our suggested IoT kit. Sections 5 provides a conclusion and future work of the research.

## 2. Literature Review

The section describes the literature review from 2017 to 2024 for our research. We have analyzed the previous studies based on IoT technology implemented for cardiac healthcare monitoring purposes. Table 1 provides a review of the literature studies done so far. The detailed assessment of the few studies is discussed below:

The research study in 2017 implemented an IoT kit for cardiac patients, which also contains an effective alarm [13]. Sensory data from the patient's body via heart rate and temperature sensor is transferred to a microcontroller platform and displayed on a Liquid Crystal Display (LCD) screen. The Wireless Fidelity (Wi-Fi) module ESP8266 communicates the sensory data to the internet/cloud. Global System for Mobile Communication (GSM) and phone calls or text messages send alerts to the patient's phone number and notify the patient's family and friends. Another research study in 2017 [14] proposed a system for heartbeat sensing and heart attack detection using IoT technology. The pulse sensor is used for heartbeat monitoring and heart attack detection systems.

The Wireless Body Sensor Network (WBSN) presented in this system was used to monitor cardiac patients' health, utilizing a three-tiered design in 2018 [15]. Data from the heart and oxygen sensors are collected from the patient's body and sent to the Thing Speak IoT application for distant site servers. Further, research publications in 2018 [16] [17] focused on detecting cardiac arrest using heartbeat or photo plethysmography (PPG) sensors. The heart rate can be tracked with the use of a PPG sensor. A low pulse rate should be communicated to the patient's doctor, hospital, or family.

This 2019 study presents an IoT-based module for diagnosing heart stroke using a heart rate and temperature sensor [18]. Placing the patient's little finger on the pulse sensor allows for detecting their heart rate. The system contains an alert system in case of a patient emergency. Another research of 2019 [19] suggests developing a wearable IoT solution for cardiac patients that can measure their body temperature and heart rate directly from their bodies using a machine learning technique. A temperature sensor and pulse sensor data are sent to an Arduino UNO over Bluetooth to connect with a mobile device. Data from mobile applications may also be utilized to issue an emergency notice and monitor a patient's status. The author's research objective [20] is to diagnose myocardial infarction using a web application. This suggested system makes use of an electrocardiogram (ECG) sensor. The authors discuss better using hardware and software to build a web app that connects to a Google Firebase database. The app may then show the patient's electrocardiogram signal that has come in.

The research detailed in 2020 [21] developed a heart attack detection system using sensory modules. Dispersed sensors like temperature and blood pressure are used to build an IoT kit. The system's main objective is to provide a convenient and accurate way of monitoring remote patient health using an online application. A few heart rate measurements were obtained using the sensors incorporated in the design. A pulse rate sensor is used to assess the patient's heartbeat monitoring. It can use Bluetooth to transfer sensor data from a mobile device to an online application. An online application contains the patient's medical history and present condition. Another suggested 2020 [22] solution is a distributed IoT kit for health monitoring made explicitly for elderly patients. The MPX10 series pressure sensor, LM35 series temperature sensor, pulse sensor, and ECG module AD8232 are all included in the recommended IoT kit. The sensory data from the patient's body is sent to the application for monitoring. Moreover, in 2020 [23], research was conducted to develop an IoT-based system for heart rate and alcohol detection in intelligent transportation systems. Another detailed 2020 research [24] presents three-lead VCG signals to automatically identify individuals with Posterior myocardial infarction (PMI) from healthy control (HC) participants.

In the research project 2021 [25], an IoT technology consisting of an LM35 temperature sensor, digital sphygmomanometer, heartbeats sensor, and Pulse oximeters for intelligent healthcare monitoring. Moreover, the studies presented in 2021 [26] [27] [28] use only ECG signals for the detection of cardiac arrest.

The study published in 2022 [29] enhances the patient's quality of life by providing real-time insight into their status using physiological parameter measurements, such as systolic and diastolic blood pressure and body temperature. The central concept is to provide patients with care by continuously monitoring their body temperature, blood pressure, and pulse rate without requiring them to switch between facilities so that their health is constantly monitored. The temperature and blood pressure sensors' collected data is then processed and saved on the cloud so that the patient's caretakers can see it from anywhere and act accordingly in response to an alarm. Another publication of 2022 [30] developed a cloud-based healthcare monitoring system for cardiovascular patients using Bi-LSTM (bidirectional long short-term memory). The system monitors cardiac readings such as BP, heart rate, and ECG and applies machine-learning techniques. The study was presented in 2020 [31] to deliver an ECG monitoring system for healthcare users in the smart city context.

The research published in 2023 [32] uses machine learning and Internet of Things-based health monitoring systems for cardiac patients. The Internet of Things technology is used to track and report on the activities of cardiac patients, such as heart rate, ECG, temperature, and BP. Another study of 2023 [33] uses Long Short Term Memory (LSTM) and Recurrent Neural Network (RNN) for IoT-based prediction of heart health monitoring. This system considers ECG, BP, heart rate, and oxygen level readings. The paper in 2023 [34] presents an IoT-based application analysis of the patient's systolic, diastolic, and heart rates using a variety of sensors that are automated to record directly to the application database for analysis. This method is based on the lean UX approach from the Gothelf and Seiden framework.

This paper, published in 2024 [35], implemented a DEEP-CARDIO, an IoT network, to provide nutrition advice, treatment suggestions, and early diagnosis for cardiac conditions. The system used ECG, pressure sensor, and pulse sensor for data acquisition from the patient body and applied machine learning for further prediction. The recommendation system offers nutritional and activity advice to cardiac patients through a user-friendly smartphone application based on the identified data. Another study of 2024 [36] worked on developing cardiac care that uses the IoT and edge cloud computing [61]. The first part of this system involves collecting data from sensors that detect vital signs like temperature, heart rate, and blood pressure. The second part consists of processing the data in the edge cloud using algorithms like Convolutional Neural Network (CNN), transfer learning, and Haar Wavelet transform. The latest research of 2024 [37] uses various sensors such as ECG Shield, Heart Rate Muscle, and DS18B20 temperature sensors to develop a healthcare monitoring system. These sensors work together to provide a whole system for real-time tracking of patients. Heart Rate Muscle Sensors offer helpful information on cardiovascular health, and the ECG Shield guarantees continuous electrocardiogram monitoring for early identification of cardiac abnormalities.

**Table 1. Detailed Review of Literature Studies**

| Year | No. of Heart Metrics Addresses | Heart Metrics Addressed | | | | | State |
|---|---|---|---|---|---|---|---|
| | | HR | BP | SPO2 | Temp | ECG | |
| **2017 [13]** | 2 | Y | N | N | Y | N | India |
| **2017 [14]** | 1 | Y | N | N | N | N | India |
| **2017 [38]** | 4 | Y | Y | Y | N | Y | China |
| **2017 [39]** | 2 | Y | Y | N | N | N | India |
| **2017 [40]** | 4 | Y | Y | N | Y | Y | India |
| **2017 [41]** | 1 | N | N | N | N | Y | USA |
| **2018 [15]** | 2 | Y | N | Y | N | N | India |
| **2018 [16]** | 1 | Y | N | N | N | N | India |
| **2018 [17]** | 1 | Y | N | N | N | N | India |
| **2018 [42]** | 2 | Y | N | N | N | Y | India |
| **2018 [43]** | 3 | Y | N | Y | N | Y | India |
| **2018 [44]** | 3 | Y | Y | N | Y | N | India |
| **2018 [45]** | 3 | N | Y | Y | N | Y | USA |
| **2019 [18]** | 2 | Y | N | N | Y | N | India |
| **2019 [19]** | 3 | Y | N | N | Y | Y | USA |
| **2019 [46]** | 2 | Y | N | N | Y | N | India |
| **2019 [20]** | 1 | N | N | N | N | Y | India |
| **2019 [47]** | 2 | Y | N | N | N | Y | USA |
| **2019 [48]** | 2 | Y | Y | N | N | N | India |
| **2019 [49]** | 3 | Y | Y | N | N | Y | Bangladesh |
| **2019 [50]** | 1 | N | N | N | N | Y | India |
| **2019 [51]** | 1 | N | N | N | N | Y | Taiwan |
| **2020 [52]** | 1 | N | N | N | N | Y | India |
| **2020 [21]** | 3 | Y | Y | N | Y | N | India |
| **2020 [22]** | 5 | Y | Y | Y | Y | Y | India |
| **2020 [23]** | 1 | Y | N | N | N | N | India |
| **2020 [24]** | 1 | N | N | N | N | Y | India |
| **2020 [53]** | 2 | Y | N | N | N | Y | Malaysia |
| **2020 [54]** | 2 | N | N | N | Y | Y | Romania |
| **2020 [55]** | 1 | N | N | N | N | Y | Australia |
| **2020 [56]** | 1 | N | N | N | N | Y | Netherlands |
| **2020 [57]** | 2 | Y | N | N | N | Y | Iran |
| **2021 [58]** | 2 | Y | Y | N | N | N | Iran |
| **2021 [26]** | 1 | N | N | N | N | Y | India |
| **2021 [27]** | 1 | N | N | N | N | Y | Taiwan |
| **2021 [28]** | 1 | N | N | N | N | Y | China |
| **2021 [59]** | 3 | Y | N | Y | N | Y | Iran |
| **2021 [60]** | 1 | N | N | N | N | Y | India |
| **2022 [29]** | 3 | Y | N | Y | Y | N | India |
| **2022 [30]** | 3 | Y | N | Y | N | Y | India |
| **2022 [31]** | 1 | N | N | N | N | Y | Italy |
| **2023 [32]** | 4 | Y | N | Y | Y | Y | Pakistan |
| **2023 [33]** | 4 | Y | Y | Y | N | Y | India |
| **2023 [34]** | 2 | N | Y | Y | N | N | Korea |
| **2024 [35]** | 3 | Y | Y | N | N | Y | India |
| **2024 [36]** | 3 | Y | Y | N | Y | N | India |
| **2024 [37]** | 3 | Y | N | N | Y | Y | India |

BP-Blood Pressure, ECG-Electrocardiogram, HR-Heart Rate, N-No, SPO2- Oxygen Saturation, Temp-Temperature, Y-Yes

## 3. Proposed IoT-based Remote and Continuous Cardiovascular Patient Monitoring

The proposed methodology used for the development of an IoT-based cardiac healthcare system is discussed in the following steps:

1. Complete Working of the Proposed system Working- (Block diagram)
2. Circuit Designs
3. Hardware Kit Development
4. Cloud-based Mobile Application Development for Ubiquitous Access
5. Excel Streaming for Data Set Development

### *3.1. Complete Working of Proposed System - Block Diagram*

Our proposed remote cardiac healthcare monitoring system consists of MAX32664, ECG AD8232 (R, L, F), and an 18DS20 sensor connected to the patient body. MAX32664D sensor is used to measure oxygen saturation, heart rate and BP of a cardiac patients. ECG AD8232 senor utilizes to analyze the ECG while DS18B20 is for monitoring body temperature of cardiac patients. ECG module and temperature sensor are attached to the same microcontroller to transmit the data through Wi-Fi protocol to the Firebase cloud system. The values received from the patient's body via the Esp8266 microcontroller can be viewed on the user's and doctor's side [62]. A patient's health condition can be remotely and continuously monitored by the doctor or family member, depending on whether an android app is installed on either side. The Android app also contains the function of generating notifications/alerts if the patient's condition is abnormal. A Firebase cloud server sends patient healthcare data to the mobile application, saved on the Excel stream for Artificial Intelligence (AI) based classification. The block or flow diagram of our proposed system is shown in Figure 2.

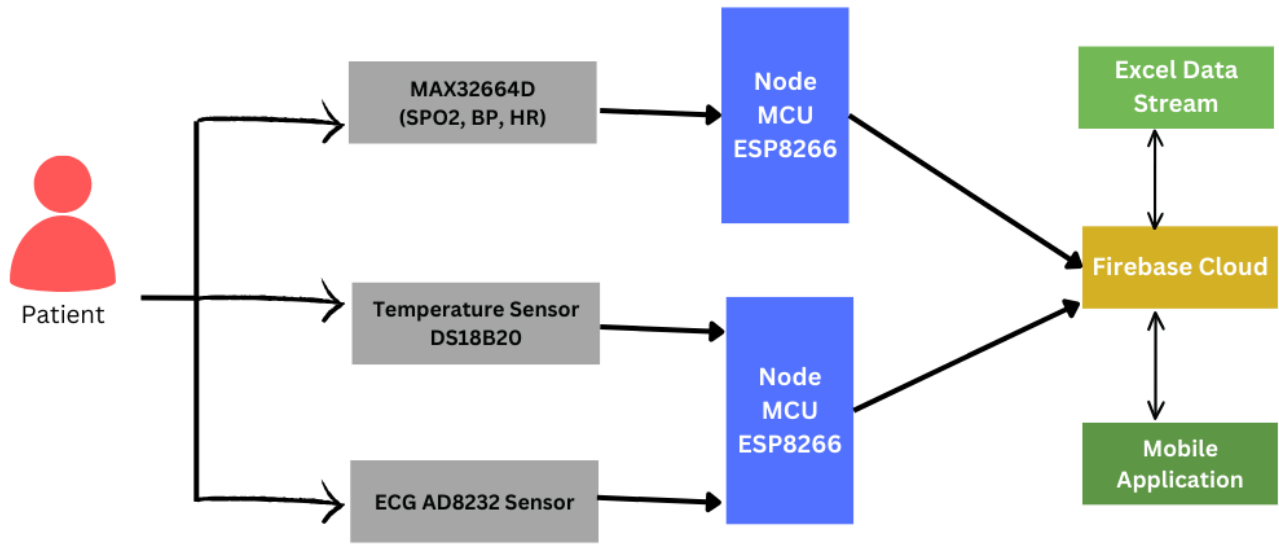

**Figure 2.** Block Diagram

*3.2. Circuit Designs*

The first step before implementing our proposed system is to design the system in a soft environment. We have developed the circuit designs for the hardware connection establishment. The circuit designs for the system are mentioned below.

3.2.1. **Circuit Design of DS18B20 Sensor**

The circuit design of the temperature sensor is shown in Figure 3. Connections for the DS18B20 sensor are mentioned below:

- GND pin attached to GND of Node MCU
- Data pin attached to D4 of Node MCU
- VCC pin attached to 3V of Node MCU

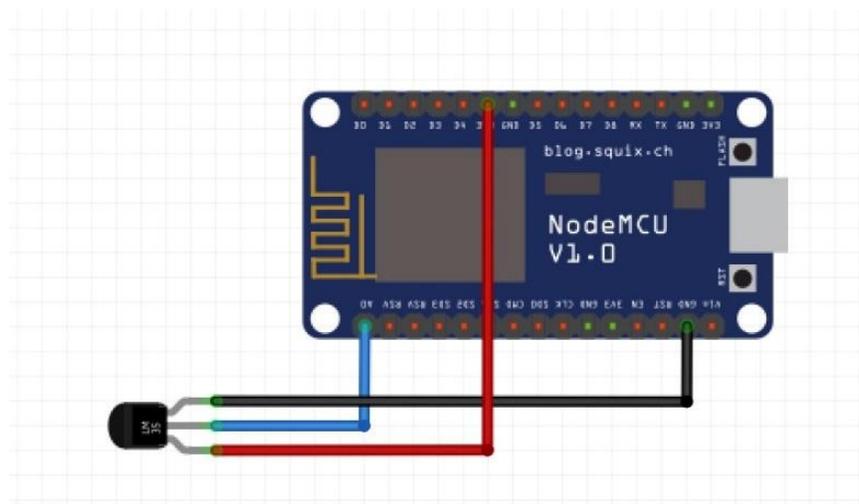

**Figure 3.** Circuit Design of DS18B20 Sensor

### 3.2.2. Circuit Design of ECG Sensor

The circuit design of the ECG AD8232 sensor is shown in Figure 4.

Connections ECG sensor are mentioned below:

- GND pin attached to GND pin of Node MCU
- 3.3V pin attached to 3V pin of Node MCU
- OUTPUT pin attached to A0 pin of Node MCU
- LO- pin attached to D7 pin of Node MCU
- LO+ pin attached to D8 pin of Node MCU

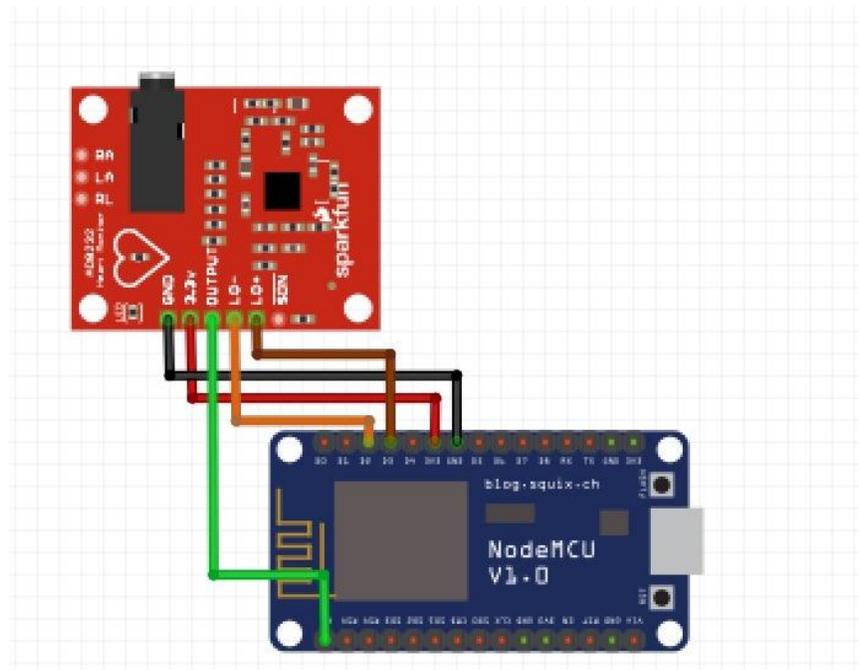

**Figure 4.** Circuit Design of ECG Sensor

3.2.3. **Circuit Design of MAX32664 Sensor**

The connections for the MAX32664D sensor are mentioned below:

- GND pin attached to GND pin of Node MCU
- RST pin attached to D4 pin of Node MCU
- MFIO pin attached to D3 pin of Node MCU
- SCL pin attached to D2 pin of Node MCU
- SDA pin attached to D1 pin of Node MCU
- VCC pin attached to 3V pin of Node MCU

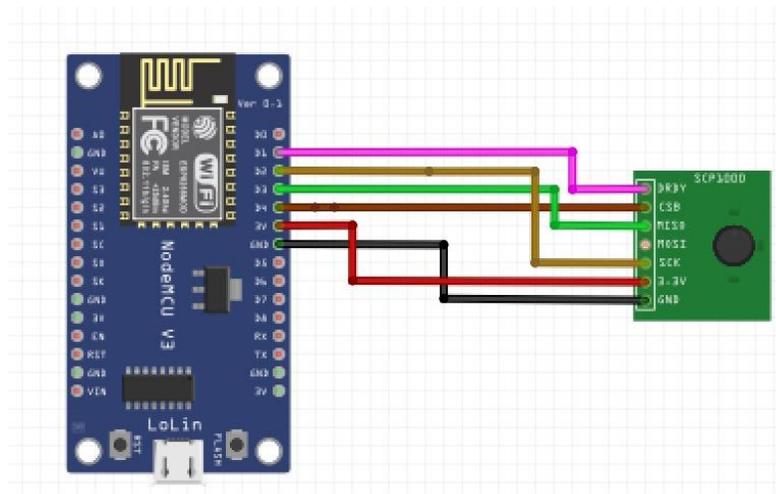

**Figure 5.** Circuit Design of MAX32664D Sensor

## 3.3. Hardware Kit Implementation

Our hardware kit consists of DS18B20, AD8232, MAX32664D, and Node MCU (Wifi module). Details of our sensors used in the hardware kit development are given below.

### 3.3.1. Temperature Sensor (DS18B20)

Maxim IC's DS18B20 is the company's most significant digital temperature sensor. Temperature series from -55C to 125C (+-0.5) while using DS18B20 has a 9 to 12-bit precision. Many temperature sensors may be connected to a single data bus since each has a unique 64-bit serial number programmed identifier. Many data logging and temperature control solutions depend on the proper temperature sensor. We have used DS18B20 for body temperature monitoring of cardiac patients [63].

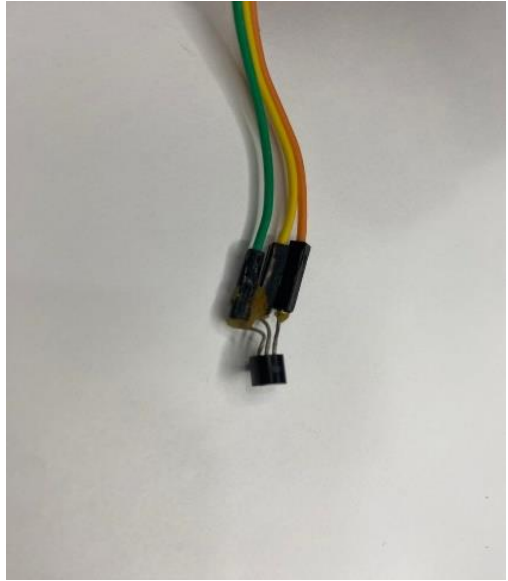

**Figure 6.** DS18B20 Sensor

### 3.3.2. ECG Sensor (AD8232)

With the single lead heart rate monitor, it's easy to keep track of your heart's electrical activity, which is a cost-effective solution. An ECG or electrocardiogram is a charted form of this electrical activity that provides an analog readout. The PR and QT intervals of the ECG can be muddled by noise, and this module functions as an op amp to assist in isolating the clear signal from the noise. ECG and other bio-potential applications benefit from this signal conditioning block. Bio-potential signals can be extracted and amplified in noisy environments, such as those caused by mobility or remote electrode placement. Analog output with an operating voltage of 3.3 volts LED indication for lead-off detection during shutdown 3.5 mm jack for connecting biomedical pads. We have used AD8232 for ECG monitoring of cardiac patients.

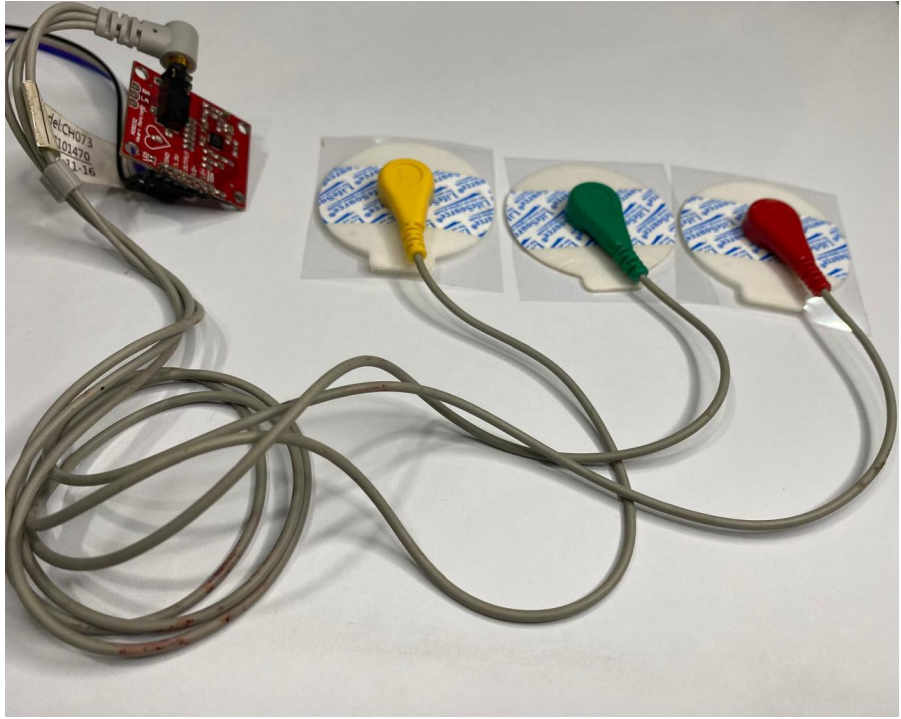

**Figure 7.** AD8232 Sensor

### 3.3.3 MAX32664 Sensor

Pulse Express has an integrated high-sensitivity optical sensor and a low-cost computation processor (biometric sensor hub MAX32664D). Pulse Express has an inherent algorithm that measures different data as you start, thanks to Maxim's MAX32664D version D integration. The board is excellent for finger-based wearable health applications because of its low-power capabilities. The data from this board may be seen using the OpenView proto central Graphical User Interface (GUI). Change the "R" value of the algorithm if SPO2 readings are unreliable or if the measured SPO2 constantly hovers at 100%. This is the case if an optical shield covers the sensor [61]. We have used the MAX322664D sensor for heart rate, oxygen saturation, and BP monitoring.

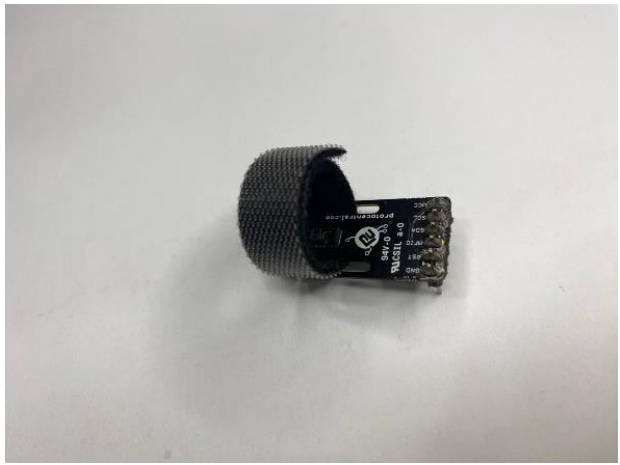

**Figure 8.** MAX32664D Sensor

### 3.4 Application Development

The Android application for the cardiac healthcare monitoring system is developed using the react native Platform.

The computer requirements are:

- 8GB RAM
- Core i3 or i5 Laptop

To develop an application, we have used a few steps given below:

Step 1: Visual Studio code on Windows 10 64-bit is being installed.

Step 2: Node JS and Node JS Package Manager (NPM) from nodejs.org need to be installed. We have installed v10.16.3.

Step 3: Setting Up React Native Development Environment.

- We must have node, react native Command Line (CLI), android studio, and Java Development Kit (JDK).
- We have used VS studio for the code editor.
- Chocolaty software is installed as a Node Package Manager (NPM).
- Choco install -y Node JS install open JDK8 (command has used to install JDK)

After that, we installed Android Studio, including:

1. Android SDK
2. Android SDK Platform
3. Android Virtual Device (for emulator).

Step 4: Application created by using React Native CLI

Step 5: Application .apk file created using the developer mode option.

Moreover, there are the following steps to generate the data from the kit:

1. Verify the device and connection establishments
2. Attach the kit to the patient's body
3. Register User on mobile application
4. Login user on the mobile application by giving credentials
5. Continuous Monitoring of patient data
6. Checking Notification and alerts (if any)
7. Save Data
8. Check saved and previous patient historical data
9. Logout

To generate the data for continuous monitoring, we must ensure that our kit is connected to the power and Wi-Fi settings. Our proposed IoT kit is shown in Figure 9.

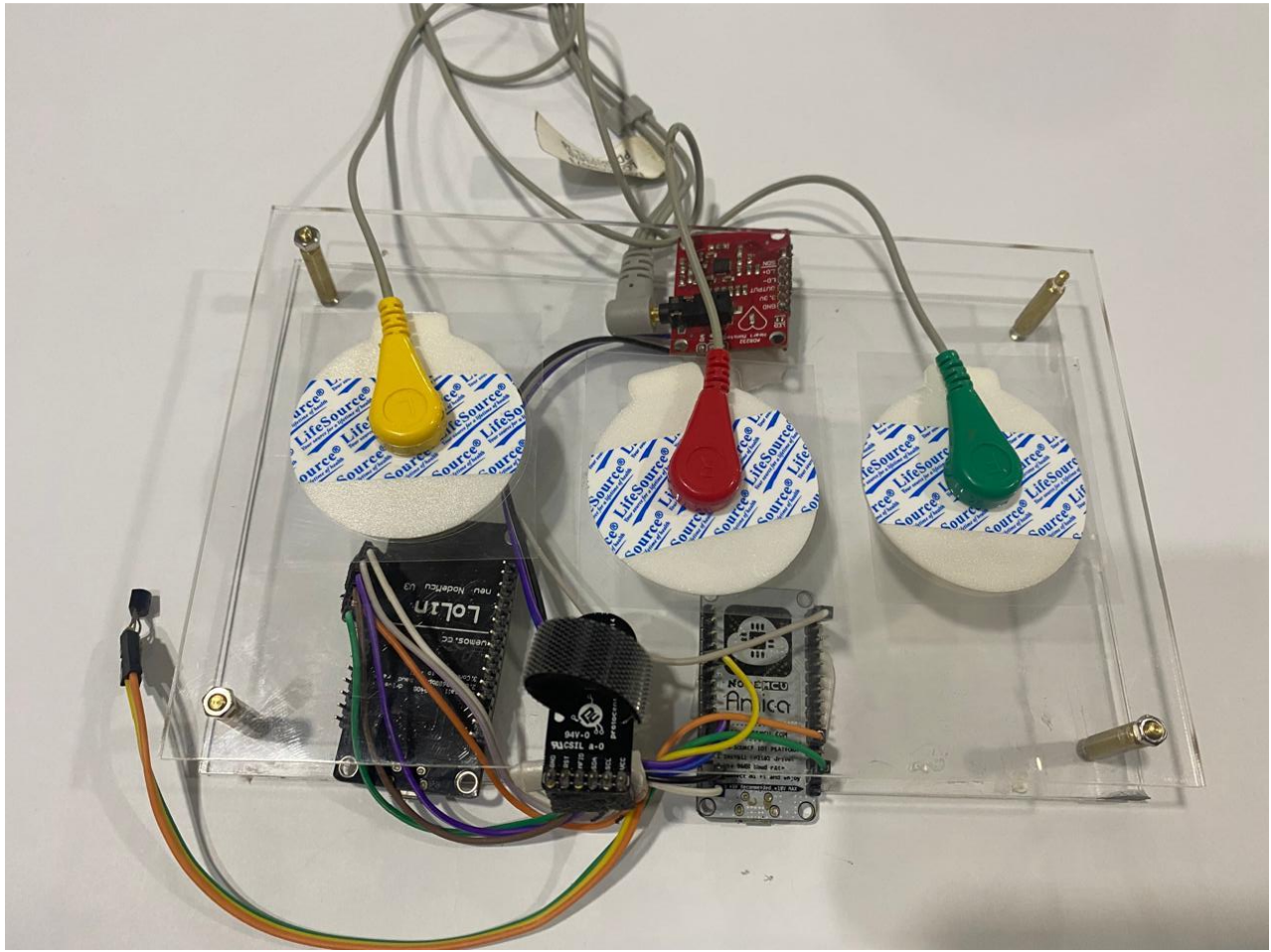

**Figure 9.** Proposed Sensor

After making sure the connections are successfully installed, we can attach the device to the patient body. The following libraries are used to establish the connection between hardware and the cloud database.

```
#include <NTPClient.h>
#include "FirebaseESP8266.h"
#include <ESP8266WiFi.h>
#include <WiFiUdp.h>
#include <TimeLib.h>
#define FIREBASE_HOST "healthcare-monitoring-sy-66936-default-rtdb.firebaseio.com/"
#define FIREBASE_AUTH "SjsrDVVR1T6uWVLebMAMh3J6eQJ5NrYvdGtdSTTM"
#define WIFI_SSID     "mywifi"
#define WIFI_PASSWORD "123456789"
```

After making sure the connections were established successfully, we allowed the patient to place a finger or attach the sensor to the body. We must register a patient before saving the data to the cloud. Providing all the necessary details such as patient name, email, contact, and password is necessary for registration as shown in the code below.

```
export default function RegisterUser({navigation}) {
  let [register, setRegister] = useState(
    {
      name: '',
      email: '',
      contact: '',
      password: '',
      c_password: ''
    });
```

User can log in with the same credentials. After logging in, the family member of a patient can see the notification alerts on the application if the patient's condition is found critical. Such as, if the patient's oxygen level decreases, then the predefined threshold, notification will continuously pop up on the application home screen. The code for providing a notification alert for oxygen saturation is mentioned below.

```
if((MAX32664.max32664Output.spo2>94)&&(MAX32664.max32664Output.spo2<100))
{
        Firebase.setString(firebaseData1,"/deviceData/Notification/Oxygen_Level","Normal");
}
if((MAX32664.max32664Output.spo2>0)&&(MAX32664.max32664Output.spo2<95))
{
        Firebase.setString(firebaseData1,"/deviceData/Notification/Oxygen_Level","AB_Normal");
}
```

On the home page of this mobile application, doctors or family members can continuously see the patient's health status by observing all basic and vital parametric values generated from the patient body. The data can be saved by clicking on the record sample button. The saved data will help the doctor to monitor the previous record of the patient's health condition.

## 3.4 Excel Stream

Excel streaming is used to receive and store the data in tabular format. The data generated from the patient's body is helpful for an expert's predictions of any abnormality. We used the Excel stream to record the ECG signals in a separate channel depending on the number of streams selected for the specific time. Our data stream for the Excel sheet consists of Data in, Data out, Settings, and Manifest. The settings used for the ECG stream are:

- A data interval of 150 ms.
- Data rows of 2000 (1.5 minutes).
- 10 data channels.
- The data orientation of the newest last.

Once the human body is connected to the L, R, and F electrodes of ECG, then the data coming from the body is recorded by this streaming file. This streaming file can be used to make further predictions about ECG data. The ECG workbook settings are shown in Table 2.

**Table 2.** ECG Workbook Settings

| ECG Workbook Parameter | ECG Workbook Parameter |
|---|---|
| Data Interval (ms) | Data Interval (ms) |
| Data Rows | Data Rows |
| Data Channels | Data Channels |
| Data Orientation | Data Orientation |

## 4. Results and Discussion

Our proposed system is tested in a natural environment. We have analyzed the system by contrasting the outcomes of our hardware kit results and data generated from the subjects' bodies via a doctor's device. Our kit has been tested on over 20 patients in the Iqram-Ul-Haq Clinic in Multan.

As mentioned, we tested our kit in a natural environment. The following results were achieved by testing the kit in a natural clinical setting on 20 patients, as shown in Table 3. We applied the IoT kit and clinical devices to the same patient. After that, we extracted the differences between the results obtained from the devices used in the doctor's clinic and the IoT Kit. The difference in the values of our IoT kit (if varied) from the doctor device is shown in Table 4.

Table 3 contains the setup name, ID, and sensor values. A Setup name consists of our proposed IoT kit and clinical setup used by medical experts. While, ID represents the patient's Identity number. The sensory data contains oxygen saturation, temperature, blood pressure, heart rate, and ECG values.

To generate the results from the patient's body we have first attached the IoT kit. Like, For the patient having ID '1' attached our proposed IoT kit contains an oxygen level of 95%, body temperature of 100°F, Systolic and Diastolic BP 120/90, heart rate with values of 70bpm, and ECG with PQRST values as 254, 450, 119, 701, 88, 76. Then the doctor's devices are attached to the same patient having ID '1' and recorded the values. After recording the values from the proposed IoT kit and clinical setup, we have compared the values. For patient '1' we have got all the similar values from the IoT kit and doctor devices as shown in

Table 4. Likewise, we have generated the data of 20 subjects, recorded them, and compared it with the expert's devices.

**Table 3.** Evaluated Kit results by comparing them with real clinical setup

| Setup Name | ID | SPo2 | Temp | S_BP | D_BP | HR | ECG | P | Q | R | S | T |
|---|---|---|---|---|---|---|---|---|---|---|---|---|
| Kit | 1 | 95 | 100 | 120 | 90 | 70 | 254 | 450 | 119 | 701 | 88 | 76 |
| Clinic | 1 | 95 | 100 | 120 | 90 | 70 | 254 | 450 | 119 | 701 | 88 | 76 |
| Kit | 2 | 96 | 101 | 125 | 95 | 73 | 253 | 449 | 115 | 669 | 78 | 75 |
| Clinic | 2 | 95 | 101 | 125 | 95 | 73 | 253 | 449 | 115 | 669 | 78 | 75 |
| Kit | 3 | 95 | 101 | 119 | 75 | 79 | 254 | 450 | 114 | 701 | 81 | 71 |
| Clinic | 3 | 95 | 101 | 119 | 75 | 79 | 254 | 450 | 114 | 701 | 81 | 71 |
| Kit | 4 | 97 | 100 | 135 | 100 | 81 | 255 | 450 | 119 | 700 | 90 | 76 |
| Clinic | 4 | 97 | 100 | 135 | 95 | 81 | 255 | 450 | 119 | 700 | 90 | 76 |
| Kit | 5 | 95 | 101 | 119 | 75 | 74 | 254 | 450 | 120 | 702 | 81 | 71 |
| Clinic | 5 | 95 | 101 | 119 | 75 | 74 | 254 | 450 | 120 | 702 | 81 | 71 |
| Kit | 6 | 97 | 100 | 135 | 100 | 80 | 255 | 460 | 119 | 700 | 91 | 74 |
| Clinic | 6 | 97 | 101 | 132 | 95 | 82 | 255 | 460 | 119 | 700 | 91 | 74 |
| Kit | 7 | 99 | 100 | 115 | 80 | 79 | 255 | 450 | 120 | 702 | 82 | 71 |
| Clinic | 7 | 99 | 100 | 115 | 80 | 79 | 255 | 450 | 120 | 702 | 82 | 71 |
| Kit | 8 | 97 | 101 | 135 | 100 | 80 | 255 | 447 | 119 | 700 | 97 | 74 |
| Clinic | 8 | 97 | 101 | 135 | 95 | 83 | 255 | 447 | 119 | 700 | 97 | 74 |
| Kit | 9 | 96 | 100 | 125 | 75 | 79 | 255 | 449 | 120 | 702 | 84 | 71 |
| Clinic | 9 | 96 | 100 | 125 | 72 | 79 | 255 | 449 | 120 | 702 | 84 | 71 |
| Kit | 10 | 97 | 101 | 135 | 100 | 72 | 255 | 447 | 119 | 700 | 97 | 78 |
| Clinic | 10 | 97 | 101 | 134 | 100 | 72 | 255 | 447 | 119 | 700 | 97 | 78 |
| Kit | 11 | 96 | 102 | 119 | 75 | 79 | 254 | 450 | 120 | 702 | 81 | 71 |
| Clinic | 11 | 95 | 101 | 116 | 75 | 79 | 254 | 450 | 120 | 702 | 81 | 71 |
| Kit | 12 | 97 | 101 | 135 | 100 | 72 | 255 | 460 | 119 | 700 | 91 | 74 |
| Clinic | 12 | 97 | 102 | 135 | 100 | 72 | 255 | 460 | 119 | 700 | 91 | 74 |
| Kit | 13 | 95 | 100 | 120 | 100 | 74 | 255 | 455 | 120 | 705 | 89 | 71 |
| Clinic | 13 | 95 | 100 | 122 | 100 | 74 | 255 | 455 | 120 | 705 | 89 | 71 |

| | | | | | | | | | | | |
|---|---|---|---|---|---|---|---|---|---|---|---|
| Kit | 14 | 96 | 102 | 135 | 100 | 72 | 255 | 447 | 119 | 700 | 99 | 74 |
| Clinic | 14 | 96 | 101 | 135 | 95 | 75 | 255 | 447 | 119 | 700 | 99 | 74 |
| Kit | 15 | 96 | 101 | 120 | 75 | 76 | 254 | 450 | 120 | 702 | 81 | 71 |
| Clinic | 15 | 95 | 101 | 120 | 75 | 76 | 254 | 450 | 120 | 702 | 81 | 71 |
| Kit | 16 | 96 | 101 | 134 | 100 | 79 | 255 | 450 | 119 | 700 | 91 | 74 |
| Clinic | 16 | 96 | 102 | 134 | 100 | 79 | 255 | 450 | 119 | 700 | 91 | 74 |
| Kit | 17 | 95 | 102 | 120 | 90 | 74 | 255 | 455 | 120 | 705 | 99 | 71 |
| Clinic | 17 | 95 | 102 | 122 | 90 | 74 | 255 | 455 | 120 | 705 | 99 | 71 |
| Kit | 18 | 97 | 101 | 135 | 100 | 80 | 255 | 447 | 119 | 706 | 82 | 73 |
| Clinic | 18 | 97 | 101 | 135 | 100 | 81 | 255 | 447 | 119 | 706 | 82 | 73 |
| Kit | 19 | 95 | 100 | 119 | 75 | 81 | 255 | 454 | 130 | 694 | 87 | 76 |
| Clinic | 19 | 95 | 100 | 119 | 73 | 81 | 255 | 454 | 130 | 694 | 87 | 76 |
| Kit | 20 | 94 | 99 | 110 | 80 | 79 | 255 | 454 | 130 | 694 | 87 | 76 |
| Clinic | 20 | 94 | 99 | 110 | 80 | 79 | 255 | 454 | 130 | 694 | 87 | 76 |

D_BP-Diastolic Blood Pressure, ECG-Electrocardiogram, S_BP-Systolic Blood Pressure, HR-Heart Rate, SpO2- Oxygen Saturation, Temp-Temperature

**Table 4.** *Difference between proposed IoT kit and Doctor Devices used in Clinical Setup.*

| ID | Spo2 | Temp | S_BP | D_BP | HR | ECG | P | Q | R | S | T |
|---|---|---|---|---|---|---|---|---|---|---|---|
| 1 | 0 | 0 | 0 | 0 | 0 | 0 | 0 | 0 | 0 | 0 | 0 |
| 2 | 1 | 0 | 0 | 0 | 0 | 0 | 0 | 0 | 0 | 0 | 0 |
| 3 | 0 | 0 | 0 | 0 | 0 | 0 | 0 | 0 | 0 | 0 | 0 |
| 4 | 0 | 0 | 0 | 5 | 0 | 0 | 0 | 0 | 0 | 0 | 0 |
| 5 | 0 | 0 | 0 | 0 | 0 | 0 | 0 | 0 | 0 | 0 | 0 |
| 6 | 0 | 1 | 3 | 5 | 2 | 0 | 0 | 0 | 0 | 0 | 0 |
| 7 | 0 | 0 | 0 | 0 | 0 | 0 | 0 | 0 | 0 | 0 | 0 |
| 8 | 0 | 0 | 0 | 5 | 3 | 0 | 0 | 0 | 0 | 0 | 0 |
| 9 | 0 | 0 | 0 | 3 | 0 | 0 | 0 | 0 | 0 | 0 | 0 |
| 10 | 0 | 0 | 1 | 0 | 0 | 0 | 0 | 0 | 0 | 0 | 0 |
| 11 | 1 | 1 | 3 | 0 | 0 | 0 | 0 | 0 | 0 | 0 | 0 |
| 12 | 0 | 1 | 0 | 0 | 0 | 0 | 0 | 0 | 0 | 0 | 0 |
| 13 | 0 | 0 | 2 | 0 | 0 | 0 | 0 | 0 | 0 | 0 | 0 |
| 14 | 0 | 1 | 0 | 5 | 3 | 0 | 0 | 0 | 0 | 0 | 0 |
| 15 | 1 | 0 | 0 | 0 | 0 | 0 | 0 | 0 | 0 | 0 | 0 |
| 16 | 0 | 1 | 0 | 0 | 0 | 0 | 0 | 0 | 0 | 0 | 0 |
| 17 | 0 | 0 | 2 | 0 | 0 | 0 | 0 | 0 | 0 | 0 | 0 |
| 18 | 0 | 0 | 0 | 0 | 1 | 0 | 0 | 0 | 0 | 0 | 0 |
| 19 | 0 | 0 | 0 | 2 | 0 | 0 | 0 | 0 | 0 | 0 | 0 |

| 20 | 0 | 0 | 0 | 0 | 0 | 0 | 0 | 0 | 0 | 0 | 0 |

D_BP-Diastolic Blood Pressure, ECG-Electrocardiogram, S_BP-Systolic Blood Pressure, HR-Heart Rate, SPO2- Oxygen Saturation, Temp-Temperature

Table 5 shows the analysis of the results of our proposed IoT kit. In this Table, we have mentioned several metrics in our remote continuous monitoring IoT kit for cardiac patients. For each metric, we stated the total readings, correctly predicted readings, incorrectly predicted readings, and nearly corrected readings. As mentioned in Table 5, the total readings for oxygen saturation are of 20 subjects collectively. From 20 readings of oxygen saturation, we got 17 subjects' values as correctly predicted readings which means that those 17 subjects out of 20 have similar readings of expert device with no difference. There is no incorrect reading for oxygen saturation. A total of 3 readings were closer to the expert's device values for oxygen saturation. Likewise, the total correctly predicted readings of body temperature were 15 out of 20 subjects with no incorrect readings and 5 subject's readings were closer to the correct values of the expert's temperature device. For the systolic BP, we got 15 correctly predicted readings out of 20 along with 5 subjects' values closer to correct readings, and got no incorrect readings. The correctly predicted values for diastolic are 14 out of 20 subjects' with zero incorrect readings and 6 subjects' values closer to correct readings of the BP device at the doctor's end. By analyzing the heart rate results, we got 16 subjects' readings correctly predicted out of 20 with no incorrect readings. However, we found 4 subjects' readings for heart rate as nearly corrected to the expert's device. And we got zero values as incorrect or nearly corrected for ECG.

**Table 5.** *Results Analysis*

| Metrics | Total Readings | Correctly Predicted Readings | Incorrectly Predicted Readings | Close To Correct Readings (Nearly Corrected) |
|---|---|---|---|---|
| Oxygen Saturation | 20 | 17 | 0 | 3 |
| Body Temperature | 20 | 15 | 0 | 5 |
| Systolic BP | 20 | 15 | 0 | 5 |
| Diastolic BP | 20 | 14 | 0 | 6 |
| Heart Rate | 20 | 16 | 0 | 4 |
| ECG | 20 | 20 | 0 | 0 |

Figure 10 below represents the results for oxygen saturation which is 17 subjects' readings were accurate or correctly predicted out of 20 for SPO2.

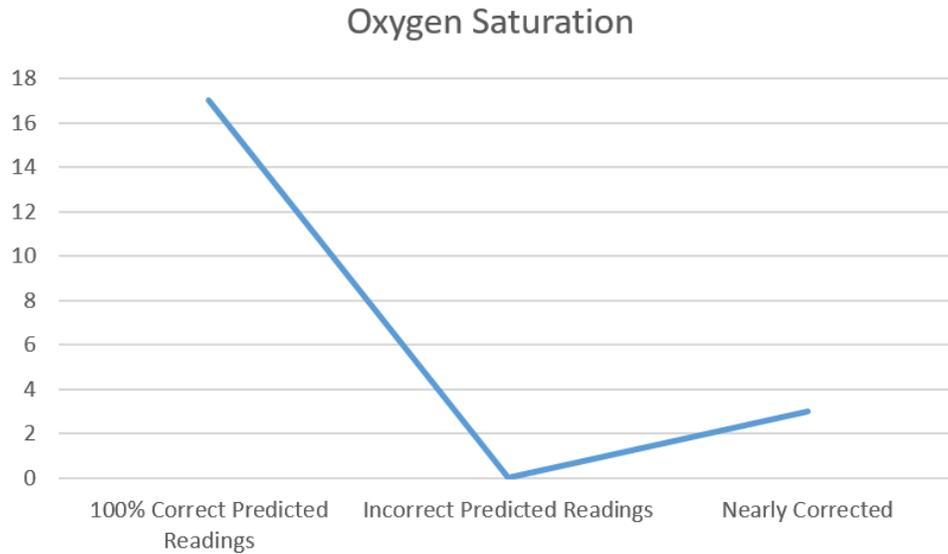

**Figure 10.** Analysis of SPO2 Readings

However, the body temperature achieves 15 subjects readings were exactly correct and the same readings out of 20 which is shown in Figure 11 below.

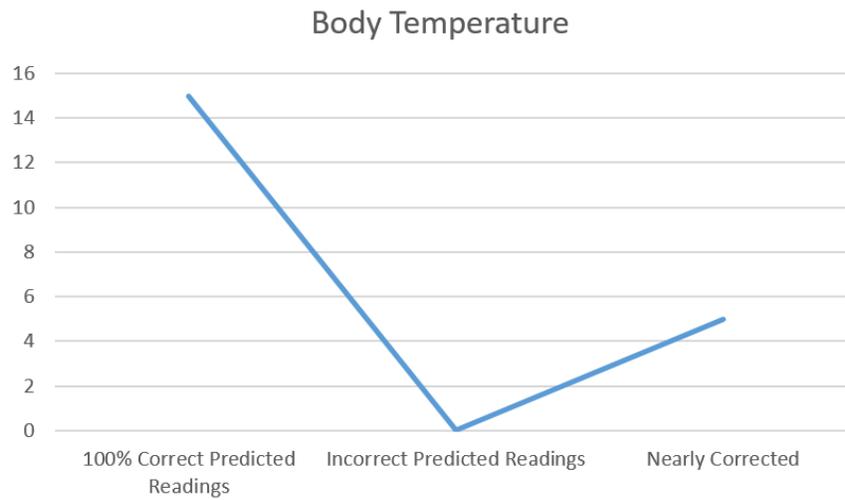

**Figure 11.** Analysis of Temperature Readings

For systolic and diastolic values, accurately anticipated readings had a ratio of 15:14 shown in Figure 12. This means systolic BP has a total of 15 subjects' readings as correctly predicted. And diastolic BP has a total of 14 subjects' readings as correctly predicted.

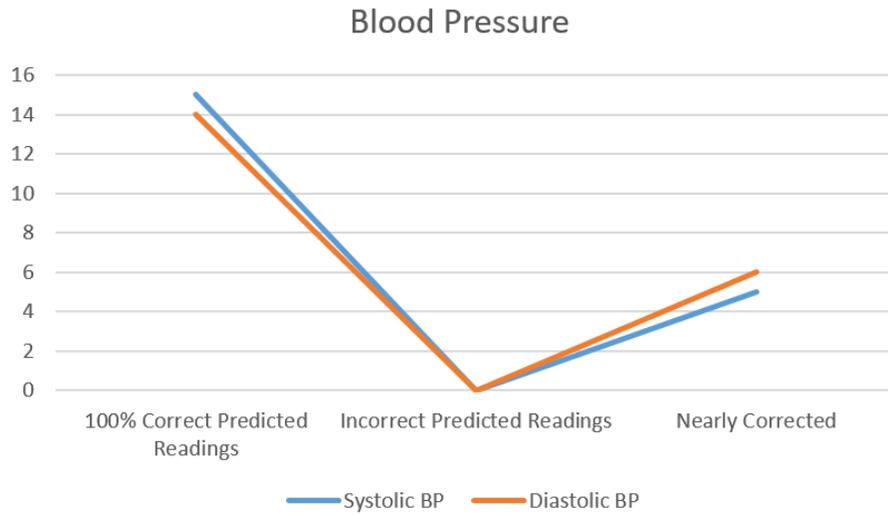

**Figure 12.** Analysis of BP Readings

However, the heart rate attains 16 subjects' readings as accurtely predicted out of 20 as shown in Figure 13 below.

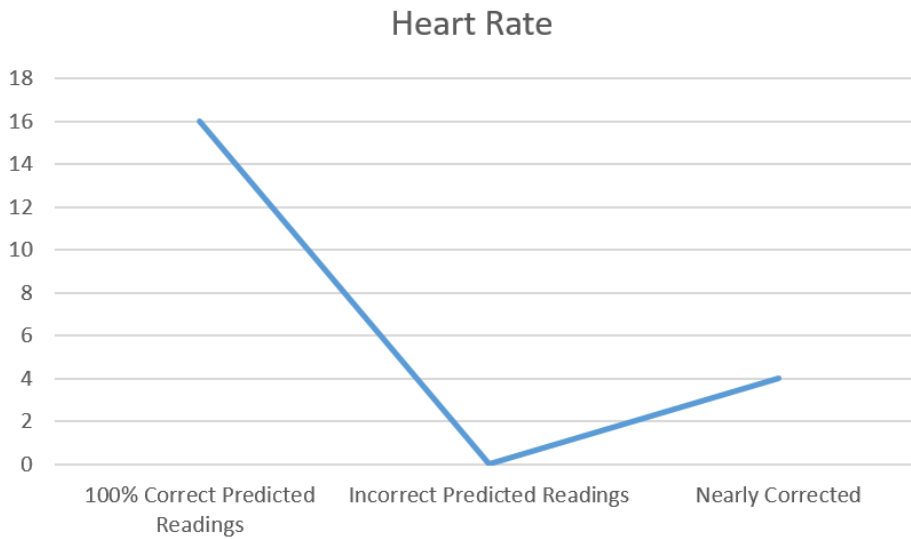

**Figure 13.** Analysis of Heart Rate Readings

The ECG reads are exactly the same with no nearly or incorrect readings shown in Figure 14.

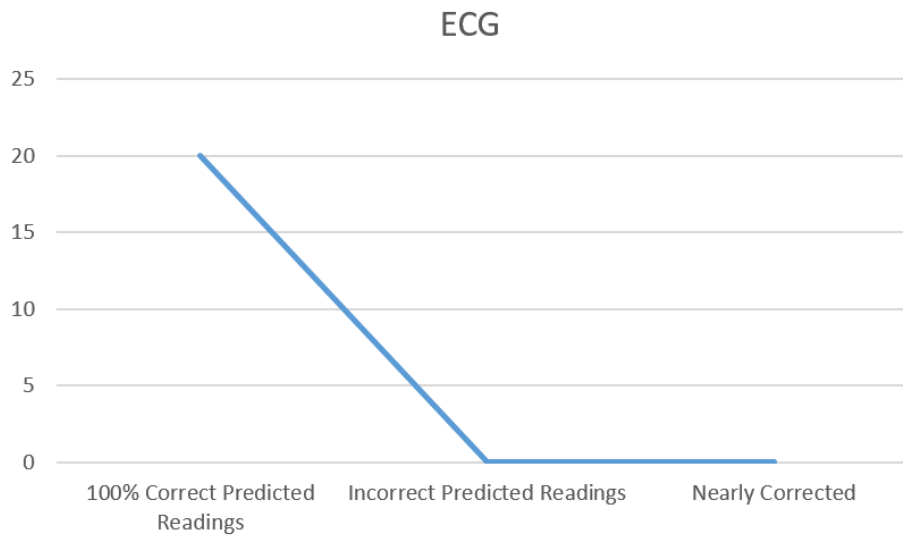

**Figure 14.** Analysis of ECG Readings

By analyzing the results, no parameter has an incorrect reading. Our sensor readings varied somewhat, even though we were able to get some nearly corrected values for all of the metrics on the 0–6 scale. According to the doctor's recommendation, these nearly corrected values are quite closer to the real levels, which is tolerable.

## 5. Conclusion and Future Work

Acute strokes and sudden deaths are significantly impacted by cardiovascular problems, especially arrhythmia, which continue to rank among the top causes of mortality worldwide. An evolution towards ubiquitous, real-time monitoring made possible by technological improvements is required to address this widespread health concern. With the help of the IoT, cardiac patients may be remotely and continuously monitored, no matter where they are, which an up-and-coming option is. This is of the utmost importance in underdeveloped countries like Pakistan, where resources and infrastructure are few.

Our research focuses on developing a cloud-based cardiovascular monitoring system that uses IoT. This system allows for continuous monitoring and quick intervention for patients with cardiovascular diseases. Several critical metrics are recorded by the system's sensing modules and sent to a cloud-based platform so that healthcare professionals may monitor the patient in real-time. The system also generates a notification if patient's condition varies from the threshold. The findings of the preliminary testing with 20 patients were encouraging and closely matched those of standard medical equipment. In the future, we want to improve this system by adding more advanced alert systems to respond quickly to emergencies.

**Funding**  The research received no external funding.

**Competing interests**  Regarding the publishing of this research, the authors state that they have no competing interests.

**Consent for publication**  Each individual authors in the study gave their informed consent.

**Data Availability Statement**  No public or private dataset has been used for this research.